\documentclass[letterpaper,twocolumn,10pt]{article}
\usepackage{usenix2019_SOUPS}

\usepackage{tikz}
\usepackage{amsmath}

\usepackage{filecontents}
\usepackage{tablefootnote}

\AtEndDocument{\par\leavevmode}

\begin{document}

\date{}

\title{\Large \bf A Key to Your Heart: Biometric Authentication Based on ECG Signals}

\def\plainauthor{Nikita Samarin}

\author{
{\rm Nikita Samarin}\\
University of California, Berkeley\thanks{Work was performed when the author was at the University of Edinburgh.}
\and
{\rm Donald Sannella}\\
University of Edinburgh
} 

\maketitle
\thecopyright

\begin{abstract}
In recent years, there has been a shift of interest towards the field of biometric authentication, which proves the identity of the user using their biological characteristics. We explore a novel biometric based on the electrical activity of the human heart in the form of electrocardiogram (ECG) signals. In order to explore the stability of ECG as a biometric, we collect data from 55 participants over two sessions with a period of 4 months in between. We also use a consumer-grade ECG monitor that is more affordable and usable than a medical-grade counterpart. Using a standard approach to evaluate our classifier, we obtain error rates of 2.4\% for data collected within one session and 9.7\% for data collected across two sessions. The experimental results suggest that ECG signals collected using a consumer-grade monitor can be successfully used for user authentication.


\end{abstract}
\section{Introduction}
Traditional passwords represent the most common mechanism of authenticating users online, despite numerous usability and security problems~\cite{adams1997making, furnell2006replacing, bonneau2012quest}. Passwords create a burden for users, as they have to be memorized and, ideally, should be long and unique. It should not come as a surprise, therefore, that many users opt to use easy-to-guess passwords that are reused across different services~\cite{ives2004domino, das2014tangled}, leading to account takeovers and personal data compromise. Research has shown, for instance, that over 50\% of users have the same passwords for different services~\cite{das2014tangled} and 81\% of data breaches occur due to poor password behavior~\cite{verizon}. 

The research community has been looking at alternative authentication schemes in order to replace or to complement traditional passwords, including push notifications~\cite{sanin2014systems}, graphical passwords~\cite{wiedenbeck2005passpoints}, trust scores~\cite{jakobsson2012implicit}, and gestures~\cite{sherman2014user}. Of particular interest is biometric authentication, which proves the identity of the user with ``something they are''. Biometrics improve system usability, as users are no longer required to remember any passwords or always carry a physical token. The ease of using a biometric for authentication has led to the rapid adoption of biometrics in private and public sectors, and the global market for biometric technology is expected to reach \$59.31 billion by 2025~\cite{grandview}.

While existing research has focused on common modalities, such as fingerprints, face recognition, and iris scans, insufficient work has been done to explore novel biometrics. In this work, we investigate a biometric based on the electrical activity of the human heart in the form of electrocardiogram (ECG) signals. Past research has demonstrated that ECG is sufficiently unique to each individual \cite{carreiras} and could be used for user authentication. This work further explores the stability (i.e. invariability) of ECG as a biometric over long periods of time. Moreover, we investigate whether ECG signals recorded using a consumer-grade ECG monitor can be used for user authentication. These monitors are more affordable and less intrusive than their medical-grade counterparts, and present a more realistic scenario of collecting an ECG from an embedded sensor. Finally, we evaluate the performance of the classifiers responsible for user authentication using two approaches, one of which is a standard method found in the existing literature and another one that provides better estimates of the mistakes made by the classifiers. 

\section{Background and Related Work}
This work evaluates ECG as a biometric for user authentication using data collected from a consumer-grade monitor over a period of four months. In this section, we provide a brief overview of biometrics and discuss related work on evaluating ECG for user authentication.

\subsection{Biometrics}
The term `biometrics' is used to describe measurable and distinctive characteristics that can be used to perform recognition of individuals~\cite{nist}. These characteristics are often divided into two categories: physiological and behavioral~\cite{nist}. Physiological biometrics relate to human physiology; these include fingerprints, facial features, iris patterns or DNA. Behavioral biometrics are based on human behavior, such as keystroke dynamics, voice or gait. In order for a biometric to be applicable for access control, it needs to be \textit{universal} (present and measurable in every individual), \textit{unique} (different in every individual), and \textit{stable} (invariant over the individual's lifetime)~\cite{handbook}. Digital representations of the unique features extracted from a biometric sample are known as \textit{biometric templates}. 

\noindent \textbf{Authentication and Identification.}
Biometrics can be used to achieve two important access control goals, user authentication and identification~\cite{handbook}. Biometric \textit{authentication} involves the user presenting an identity claim and a biometric sample. The system then decides whether this claim is valid based on the recorded biometric for this identity. In contrast, user \textit{identification} involves finding the closest match to presented biometrics among the stored templates. In this project, we focus on user authentication leaving identification as future work. 

\noindent \textbf{Limitations.}
Although biometrics offer greater usability than traditional passwords, there are still concerns over the security and privacy of biometric data~\cite{prabhakar2003biometric}. Once compromised, biometrics cannot be easily revoked, as they depend on persistent physiological or behavioral characteristics of an individual. Furthermore, operators of biometric recognition systems might obtain additional unintended information from a user's biometric data. For instance, fingerprint patterns might be correlated with certain diseases~\cite{woodward1997biometrics}. Finally, some biometric characteristics cannot be easily kept as a secret, such as an individual's face. Therefore, a user who wishes to remain anonymous might still be identified without their knowledge and consent~\cite{woodward1997biometrics}.  

\subsection{Electrocardiogram as a Biometric}
The heart is a muscle that pumps blood filled with oxygen and nutrients through the blood vessels to the body tissues \cite{heart_anatomy}. In order to pump blood, the heart muscle must contract, which generates an electrical impulse. This impulse can be detected on the surface of the body using electrodes placed on the skin, which is done during an electrocardiogram (ECG) test. An ECG trace captures the process of depolarization and repolarization of the heart chambers, which causes them to contract and relax.

Several studies examine the uniqueness and stability of ECG. Most of these follow an ``on-the-person" approach for signal acquisition, such that electrodes are located directly on the individual~\cite{biel, carreiras, cross_correlation, autoencoder, israel, russian}. There are fewer studies that follow an ``off-the-person" approach, but they illustrate a more realistic use-case scenario for ECG-based recognition systems. Such examples include installing ECG sensors into a smartphone case \cite{mobile}, embedding the sensor into a keyboard wrist rest \cite{finger_ecg}, and installing an ECG monitor into the steering wheel of a car \cite{cardiowheel}. Furthermore, existing studies often use data collected over a single data collection session as seen in~\cite{biel, mobile, israel, kyoso, string_matching, autoencoder, toronto}. Although single-session datasets are easier to create, they cannot be used to draw conclusions about the stability of ECG as a biometric. While several authors used longitudinal ECG data in their studies \cite{russian, cross_correlation}, only one study explicitly provided a side-by-side comparison of results achieved using both single-session and multiple-session data collected over a period of four months \cite{finger_ecg}. It concluded that ECG-based biometrics exhibit promising recognition rates using both short-term and long-term data. In terms of scale, most works that explore ECG for personal identification do not assess the performance of their ECG authentication systems on very large datasets, as was done for other biometric modalities. A notable exception is \cite{carreiras}, which evaluated the performance of a biometric system using a database of ECG recordings collected from 618 subjects and obtained high recognition rates. 

Using ECG for authentication can also address some of the common limitations of other biometrics. For instance, ECG cannot be observed without using dedicated sensors and, thus, can be used to make the authentication process inconspicuous. This can be useful to prevent leakage attacks, such as when an adversary obtains user credentials by shoulder surfing their victims. 
\section{Methodology}
In this section, we describe the methods used to collect and preprocess the dataset, train the classifiers to match biometric templates with presented identities, and perform the evaluation of obtained models.\footnote{A more in-depth overview of the methodology is available in the full report at \url{https://groups.inf.ed.ac.uk/tulips/projects/1718/samarin.pdf}} 

\subsection{Dataset}

We collected ECG readings from 55 participants over two sessions with a period of four months in between. Most of the subjects were affiliated with the university, either as students, support staff or faculty members. According to the demographic survey, 30 males and 25 females enrolled in the data collection aged between 18 and 60 (median $=$ 22). Furthermore, none of the participants reported any serious health issues, though several were feeling exhausted or sleep deprived at the moment of the experiment. There were no restrictions on eligibility, as long as the subject was at least 18 years old. We note that due to technical issues that occurred during data collection, only 49 participants had sufficient data points for use in subsequent analysis. 

\begin{figure}[th]
\includegraphics[width=4.5cm]{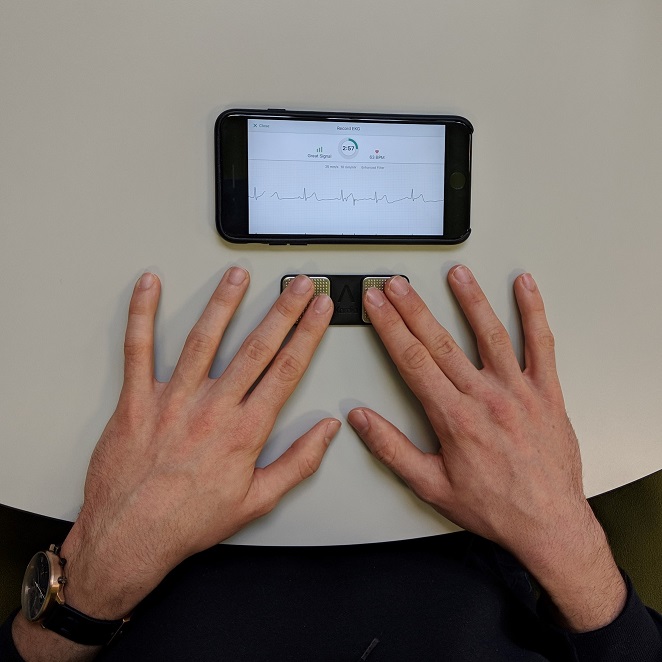}
\centering
\caption{ECG monitor connected to the smartphone application.}
\label{fig:experiment}
\end{figure}

We used an AliveCor Kardia Mobile ECG monitor~\cite{alivecor} as the best approximation to a biometric sensor that could be deployed in a real authentication system. During operation, the monitor is connected to a smartphone application, which stores the data as a single-lead ECG recording. In order to record an ECG, the user has to place two fingers from each hand onto each of the two electrodes, as shown in Figure~\ref{fig:experiment}.

\noindent \textbf{Procedure.} The data collection took place in one of the open workspaces in the School of Informatics at the University of Edinburgh. The procedure involved each participant recording their ECG trace using the monitor for 4 minutes. The recording was performed twice for a total of 8 minutes, with a break in between. The participants were not restricted in their actions and were allowed to talk and to perform movements, as long as that did not interfere with data collection. As part of a survey, subjects were asked to self-report their physical and emotional states, although this did not have any impact on the data collection. Participants who took part in both sessions received a \pounds5 Starbucks gift card at the end of the second session.

\noindent \textbf{Ethics.} The experimental methodology used in this project adheres to the ethics regulations of the University of Edinburgh and the setup was reviewed and authorized by the School of Informatics Ethics Panel. All subjects signed a consent form, which confirmed their voluntary participation in the data collection procedure.

\subsection{Data Preprocessing and Classification}
We describe the approach we took to preprocess our dataset, extract biometric templates and train classifiers. 

\begin{figure}[htbp]
\includegraphics[width=8.5cm]{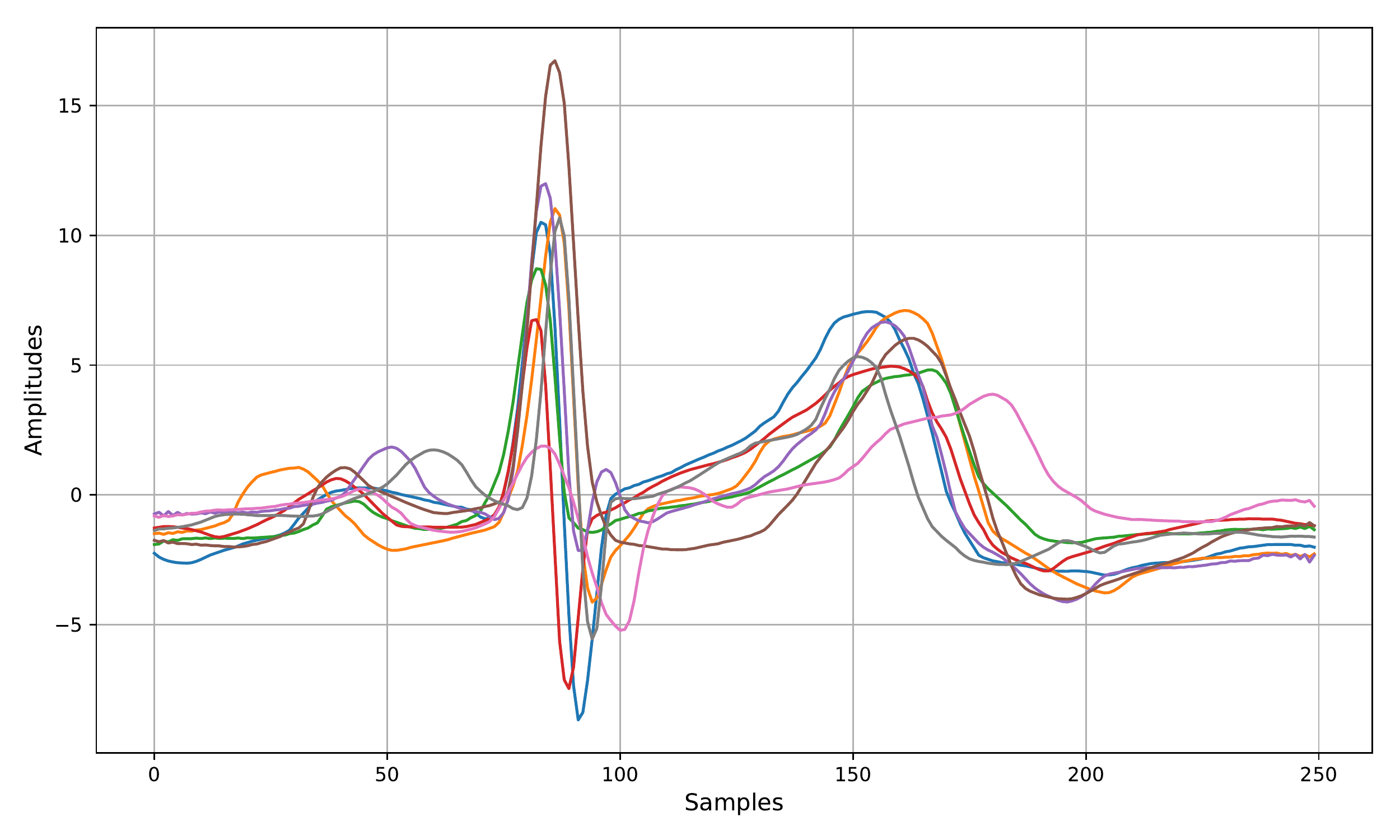}
\centering
\caption{ECG variation among 8 individuals.}
\label{fig:waveforms}
\end{figure}

\begin{figure*}[htbp]
\includegraphics[width=16cm]{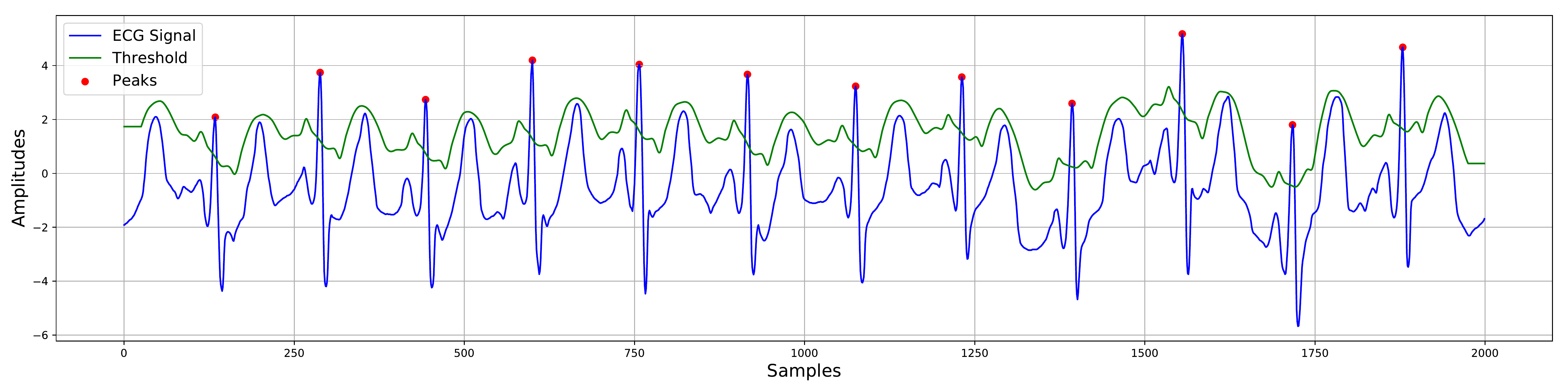}
\centering
\caption{Peak detection using a threshold based on the running mean.}
\label{fig:preprocessing}
\end{figure*}

\noindent \textbf{Signal Preprocessing.}
We preprocessed raw ECG traces before using them for evaluation. We used filters supplied by the monitor, including a Mains filter to remove power line interference and Butterworth band-pass filters to remove baseline wander noise and high-frequency noise from the ECG signal. We then divided the continuous ECG traces into segments representing individual heartbeats. First, we accentuated QRS complexes using wavelet transform and located the R peaks present in every heartbeat using a thresholding technique based on the running mean, as shown in Figure~\ref{fig:preprocessing}. We then used the located R peaks to partition the ECG traces into individual heartbeat waveforms. We removed noisy or incorrectly partitioned waveforms by comparing each segment to a median waveform of an individual and dropping 20\% of the most dissimilar segments as defined by the Euclidean distance. We used the remaining heartbeat waveforms as features to train the classifiers for user authentication. We additionally standardized the features using z-scores and performed Principal Component Analysis (PCA) to select the first 25 principal components as the input features. Figure~\ref{fig:waveforms} illustrates the obtained heartbeat waveforms (before standardization) and demonstrates the intersubject variability of the ECG. 

\noindent \textbf{Template Classification.}
In order to match biometric templates with the provided identity, we experimented with several machine learning algorithms, including logistic regression, k-nearest neighbors, and support vector machines (SVM). We chose SVM as our final model and performed 5-fold cross-validation to select the hyperparameters for the model using 80\% of data as the training set, leaving the remaining 20\% as the test set. It is important to note that we trained separate models for each user in our dataset, such that each classifier aims to predict the probability with which a provided biometric template belongs to that specific user.

\subsection{Evaluation}
In general, a biometric system can exhibit two types of errors. A \textit{false accept} happens whenever a system incorrectly accepts an intruder and a \textit{false reject} happens whenever a system incorrectly rejects a genuine user. The decision threshold of the classifier can be further tuned to improve either the overall usability (reduce the number of false rejects) or security (reduce the number of false accepts) of the system~\cite{learning_biometrics}. As is common in biometrics research~\cite{eberz}, we used equal error rate (ERR) and half target error rate (HTER) as our performance metrics. ERR is the error rate that is achieved when the decision threshold of the classifier is tuned such that the number of false rejects and false accepts is equal, while HTER represents the error rate at some predefined decision threshold. 

In this work, we evaluated the performance of our authentication models using two different approaches. Using the first method, we included the same users in the training and test sets for each evaluated classifier. While this approach is easier and commonly seen in the literature~\cite{eberz}, it underestimates the number of false accepts, as the classifier learns to distinguish the target user (for whom the classifier is trained) from every other user in the dataset. The second approach is to exclude data of a specific (non-target) user from the training set, but retain it in the test set. Therefore, during the training phase, the classifier does not learn to distinguish the readings from the target user from the readings of the excluded user, minimizing the bias in the evaluation. For each evaluated classifier, we can repeat this procedure excluding a different user every time. 



\section{Results}
We present the results of our evaluation using the two discussed approaches. For each dataset collected during the two data collection sessions, we used the first 80\% of the ECG trace across 49 subjects as the training data and the remaining 20\% of the signal as the test data. Thus, we obtained two training sets and two test sets from each session in total. 

We are also interested in examining the stability of ECG as a biometric, in other words, how invariant it remains for each individual over long periods of time. For this reason, each table includes results under three conditions. In the first two, both training and test sets are taken from the same data collection session (first or second). In the third condition, the training set is taken from the first session, while the test set comes from the second session, collected four months later. 

The results of the first evaluation approach are presented in Table~\ref{tab:first}. The average performance of classifiers for each target user is assessed using equal error rate (EER) as the metric, presented as the average of individual EER scores obtained for each of the 49 users.   

\begin{table}[htbp]
\centering
  \begin{tabular}{c c c c}
  \hline
   \hline
    Training & Testing & Average EER & Standard Deviation\\
    \hline
    S1 & S1 & 3.22\% & 2.99\% \\
    S2 & S2 & \textbf{2.44\%} & 2.40\% \\
    S1 & S2 & 9.65\% & 11.35\% \\
    \hline
    \hline
  \end{tabular}
\caption{Results obtained using the first evaluation approach. The first two columns reflect from which session (S1 or S2) the corresponding dataset originates. Lower scores indicate better performance.}
\label{tab:first}
\end{table}

The results of the second evaluation approach are shown in Table~\ref{tab:second}. In this case, we use half target error rate (HTER) as the metric, as we set the decision threshold during the training process, which represents a more realistic scenario. We obtain the average HTER score by averaging the individual HTER scores for each evaluated classifier.  

\begin{table}[htbp]
\centering
  \begin{tabular}{c c c c}
  \hline
   \hline
    Training & Testing & Average HTER & Standard Deviation\\
    \hline
    S1 & S1 & 5.86\% & 10.00\% \\
    S2 & S2 & \textbf{4.58\%} & 9.35\% \\
    S1 & S2 & 30.02\% & 17.40\% \\
    \hline
    \hline
  \end{tabular}
\caption{Results obtained using the second evaluation approach. The first two columns reflect from which session (S1 or S2) the corresponding dataset originates. Lower scores indicate better performance.}
\label{tab:second}
\end{table}

\section{Discussion and Conclusion}
In this work, we examine the usage of ECG as a biometric, focusing on the stability of the ECG signal and performance of classifiers trained using data collected from a consumer-grade ECG monitor. Comparing results reported in the literature proves to be difficult in practice, as no standardized dataset for ECG-based biometric research exists, and different authors collect their data under different conditions. Nevertheless, we present our results alongside existing studies in Table~\ref{tab:results}. We only list studies that follow a more realistic and usable ``off-the-person`` approach, in which the monitor sensors are not placed directly on the individual.  

\begin{table}[htbp]
\vspace{3mm} 
\centering
  \begin{tabular}{l c c c c c}
  \hline
   \hline
    Study & Subjects & Duration & EER \\
    \hline
    \textbf{Present Work} & 49 & Short & \textbf{2.4\%}  \\
    \textbf{Present Work} & 49 & Long  & \textbf{9.7\%}   \\
    Carreiras et al.~\cite{carreiras} & 63 & Short & \textbf{13.3\%}   \\
    Coutinho et al~\cite{string_matching} & 19 & Short & \textbf{0.4\%} \\
    Falconi et al.\tablefootnote{Authors do not provide EER, thus HTER is presented instead.}~\cite{mobile} & 10 & Short & \textbf{9.8\%} \\
    Silva et al.~\cite{finger_ecg} & 63 & Short & \textbf{1.0\%}  \\
    Silva et al.~\cite{finger_ecg} & 63 & Long & \textbf{9.1\%}  \\
    Singh et al.~\cite{singh} & 126 & Short & \textbf{3.4\%}  \\
    Komeili et al.~\cite{toronto}  & 70 & Short & \textbf{11.0\%}  \\
    \hline
    \hline
  \end{tabular}
\caption{Results from studies on ECG-based biometric authentication. All studies follow the ``off-the-person'' approach and use a single-lead ECG monitor. `Duration' indicates whether the result is obtained using short- or long-term data.}
\label{tab:results}
\end{table}

The results presented in this work provide a positive perspective on ECG-based biometrics, by showing that individuals can be authenticated by using their ECG trace. This project has also confirmed the results of previous authors showing that the performance of ECG biometrics degrades over time. Improving the performance of ECG over longer periods of time could be done by synchronizing the stored biometric with the new signal after each successful authentication.

This work also demonstrates a high potential of using consumer-grade ECG monitors for authentication. The introduction of low-cost sensors allows system designers to embed them into existing access control systems. Nevertheless, more research needs to be done on extracting features from ECG signals obtained from consumer-grade monitors, preventing spoofing attacks and guaranteeing that ECG-based biometric systems are socially accepted by the general public.


\bibliographystyle{plain}
\bibliography{references}

\end{document}